# Lessons learned from the TMT site testing campaign

T. Travouillon*[a], S.G. Els[b], R.L. Riddle[c], M. Schöck[a], A.W. Skidmore[a],

[a]TMT Observatory Corporation, 2632 E. Washington Blvd, Pasadena, CA 91107, USA
[b]European Space Astronomy Ctr, P.O. Box, 78, 28691 Villanueva de la Cañada, Madrid, Spain
[c]California Institute of Technology, 1200 E. California Blvd, Pasadena, CA 91125, USA

## ABSTRACT

After a site testing campaign spanning 5 sites over a period of 5 years, the site selection for the Thirty Meter Telescope (TMT) culminated with the choice of Mauna Kea 13N in Hawaii. During the campaign, a lot practical lessons were learned by our team and these lessons can be shared with current and future site testing campaign done for other observatories. These lessons apply to the pre-selection of the site, the ground work and operations of the campaign as well as the analysis of the data. We present of selection of such lessons in this paper preceded by a short summary of the TMT site testing activities.

## INTRODUCTION

The Thirty Meter Telescope is one of the next generation optical telescopes scheduled for the end of the decade. The telescope, which consists of 492 segments to fill a total aperture equivalent to a 30m mirror, is designed from the start to work with adaptive optics in order to take full advantage of its resolution capabilities. To make this possible, the observatory must be built on a site where the atmospheric conditions have the lowest impact. To find such a location, a site testing campaign was started in 2001 [1]. A study based on satellite observations and known conditions on the ground led to the pre-selection of 5 sites to be tested. Three sites are located in Northern Chile: Cerro Tolar, Armazones and Tolonchar, and two sites are in the northern hemisphere: San Pedro Martir in Mexico and Mauna Kea 13N in Hawaii. The onsite work started in 2004 and was concluded with the selection of Mauna Kea 13N in 2009. The goal of the site testing campaign was to gather a statistically significant set of data (a minimum of 2 years of data at each site) of the typical astro-meteorological parameters with special attention given to comparability. The parameters measured include seeing, turbulence profiles, weather parameters, precipitable water vapor, cloud cover and dust levels. The data is now available publically on the TMT site testing database (sitedata.tmt.org). The data being freely available to the community, we focus in this paper on the practical lessons learned during the site testing campaign. They consist of some of the choices we made and would recommend to other group working on site testing; but also of the things we learned ourselves along the way or even did wrong. We present these lessons organized in the different aspects of the campaign work starting from the pre-selection and finishing with the data analysis.

# EARLY CHOICES

Site testing started with the commission of a satellite study centered on cloud cover and precipitable water vapor (PWV). The goal of this study was to narrow down the search to 5 sites that would be subject to in-situ testing. This pre-selection is important for two reasons. First, all five sites turned out to be good sites; and no time and resources were therefore wasted testing poor sites thanks to the effectiveness of the pre-selection. More importantly, the satellite study gives access to a dataset spanning a much longer time period that could be accessed on the ground. In the case of cloud cover and PWV, the data collected on the ground with the site testing equipment allowed us to assess the accuracy of the satellite data over the same time period. Once validated, the satellite was used to look at longer temporal trends, and verify the statistical representativeness of the period of time which was studied from the ground.

A strong pre-selection campaign is not the only way to maximize the efficiency of the onsite campaign. The "Buy, don't build" philosophy of big science projects is a good way to avoid the trap of turning a site testing campaign into an engineering project. The focus should be to spend time gathering data and not designing instruments. Purchasing commercially available instruments (e.g. Sodar, telescopes) or using proven instruments designed and accepted by the community (e.g. MASS) allowed TMT to focus on the operation of the instrumentation and on quality control. Avoiding spending time on instrument design and fabrication allowed us to maximize the amount of data gathered at each site while at the same time quantify the errors and reproducibility of each instrument, in effect allowing us to make comparable measurements on all sites.

The importance of making quantitatively comparable measurements also led the need of standard and redundant instrumentation. Standardization is easy in the sense that it can be accomplished by using the same suite of instruments on each site. This was done for all instruments with the exception of Sodars and IRMA which we didn't have in enough quantity to populate all sites simultaneously. In these two cases, the instruments were rotated amongst the sites. For all other instruments, spare units were purchased to minimize downtime in case of malfunction. A different kind of redundancy was also instituted early in the campaign: the redundancy of expertise. In a small core group of 5 scientists, the expertise of each instrument had to be shared by two people. This was important for the analysis of the data which is mentioned later in this paper but also to prevent single point failures due to the potential inactivity of one of the group member; in other words: no one was made essential to the success of the campaign.

Making comparable measurements also requires having a muti-year campaign that is also simultaneous at all sites. Measuring over several years helps avoid or at least identify atypical years or seasons and simply improves the statistical representativeness of the measurements. Using Chile as an example, an event like the Bolivian winter can strongly affect the astro-weather, giving very different conditions than during more regular weather patterns. Measuring at all sites during the same period also helps identify the extent of such local phenomenon and avoids biases caused by global trends . In the case of the TMT site testing where 3 of the 5 sites studied are in Northern

Chile, having simultaneous measurements allowed to "map" such patterns and understand better their overall process.

## CALIBRATION CAMPAIGNS

As mentioned above, in a multiple site campaign, comparability is essential. To achieve it at an acceptable level of accuracy, standardization of the instrumentation is not sufficient. This is particularly true of parameters like atmospheric turbulence which are very localized and fast changing [2]. It is hence essential to cross calibrate each type of instruments to access accuracy and identify potential causes of offsets. The calibration campaign of the MASS/DIMM instrument, to use it as an example, was done on Cerro Tololo during a campaign lasting several weeks. Two MASS/DIMMs were run a few feet from each other during that time accumulating seeing and turbulence measurements. In addition to providing the level of reproducibility, this campaign helped identify a source of error caused by imperfections of the telescope optical alignment. A portable DIMM was also used during this calibration campaign and subsequently carried around the sites to verify the accuracy of the rest of the MASS/DIMMs. Naturally, a lot of lessons were learned from this calibration run involving complex instrumentations and measurements. What was less expected, was to also learn a lot from calibration of simpler and more commercial instruments. A calibration campaign of a few days was also carried out on Armazones using the weather stations. During this campaign, the gaps were found in the range of the humidity sensors with many instruments not reporting values between the 20 to 30% humidity range. The overall lesson that should be conveyed here, is that every parameter measured by every instrument should be verified and that practical aspects of the measurements often affect the quoted accuracy of these instruments.

## HARDWARE CHOICES

Hardware choices are very important when considering an automated site testing campaign over several sites. Seeing and turbulence profiling being of the upmost importance in this campaign, the choice of telescope used to host the MASS/DIMM units used by TMT was primordial. The 35 cm telescope used for the TMT site testing was custom built by Halfmann. It is much more rugged than the typical Meade telescope used in site testing and therefore better suited to robotic operations. It is mounted on a 6 m tower designed for the ATST site testing which is stiff enough to keep vibrations to a minimum even in high wind conditions. The telescope itself is also heavy was operated in wind speeds up to 60 km/h. This is an important feature as seeing was shown to have a dependence on wind speed, as shown in [3]. Operating in high winds therefore minimizes the bias towards lower seeing values that less robust DIMM telescopes will experience. Another important characteristic of this telescope is its open tube and open dome design. During operation, the primary mirror is directly exposed to the ambient wind, with a minimum cross section that could otherwise create locally induced turbulence. This was verified during a subsequent campaign on Cerro Tololo against a differently designed DIMM. As a whole, and considering how standard seeing and turbulence profiles have become in site testing, the instrument hosting telescope is worth considerable investment and should be considered carefully.

Another instrument that deserves to be mentioned and recommended is the sonic-anemometer. This instrument has emerged in the last decade as a better alternative to measure wind velocities. Due to the altitude of typical astronomy sites, wind vanes freeze solid in very low temperature conditions. In robotic operation it then becomes difficult to differentiate a true zero wind speed from one due to a frozen sensor. Sonic-anemometers have no moving parts and therefore do not have this problem. In addition, their cadence (several tens of Hz) allow the access to the wind power spectrum and combined with their ability to also measure temperature at the same rate gives the possibility to measure optical turbulence. These characteristics make them ideal for astronomical site testing and will hopefully become more affordable.

## SOFTWARE AND ANALYSIS

Software management was also crucial to the success of the robotic campaign [4]. Each site ran fully automated, requiring no human intervention for operation. A supervising computer managed the operation of each instrument and the communication to the outside world. A central server based at Caltech in Pasadena and mirrored at CTIO in Chile received the data from each site each morning. All raw data were kept in order to make reprocessing of the data possible and daily logs were also recorded. The sites were also in constant internet connection to the server and through a web interface allowed for the monitoring of the activities of the sites. This allowed for fast human intervention in case some of the instruments were having problems operating. This helped maximize the amount of data taken at each site and keep the hardware in a healthy state. In addition to its database functionality, the central server was also used for software and configuration backup through which all upgrades and variations to the operating software were applied. This helped avoid discrepancies between the sites which would have caused difficulties during the analysis of the data.

Finally, one of the most important lessons learned during the TMT site testing campaign regards the analysis of the data. To guarantee the accuracy of the statistics and plots presented in the final report of the campaign, every part of the analysis was done independently by two persons. This had the impact of highlighting different methods and ways to filter the data that lead to different results for the same quantity. Discovering these differences and making decisions on which method and data filters were more appropriate help make the overall analysis more systematic and robust. Once the analysis methods and criteria are set and the results verified, the analysis routines are consolidated and the reports written from these routine. It is worth mentioning that the analysis was started early in the campaign and once the routines were set, the results were simply updated by running the analysis over the latest set of data. This early start ended up being essential as some details of the analysis took some time to become apparent and a late start would have therefore meant that some results would not have been as well understood or trustworthy as they are now.

# CONCLUSION

The combination of lessons learned during the TMT site testing campaign may not all be applicable to more modestly size campaigns. But a lot has been learned in the operation of site testing campaigns involving multiple sites. From the scientific decisions to the more practical aspects of the project, the campaign was an overall success and will hopefully influence and help future campaigns. As telescopes become larger, so do their requirements for great site conditions and their testing campaign must therefore reflect this demand. It is therefore by standardizing both instrumentation and analysis that sites can be compared to the required level of accuracy and their atmospheric conditions properly understood.

# AKNOWLEDGEMENT

The TMT Project gratefully acknowledges the support of the TMT partner institutions. They are the Association of Canadian Universities for Research in Astronomy (ACURA), the California Institute of Technology and the University of California. This work was supported as well by the Gordon and Betty Moore Foundation, the Canada Foundation for Innovation, the Ontario Ministry of Research and Innovation, the National Research Council of Canada, the Natural Sciences and Engineering Research Council of Canada, the British Columbia Knowledge Development Fund, the Association of Universities for Research in Astronomy (AURA) and the U.S. National Science Foundation.